\documentclass[12pt]{article}
\usepackage{amsfonts}
\usepackage{amssymb}
\usepackage{graphicx}
\newcounter{rown}

\def\siz{\small}
\newcommand{\be}{\begin{equation}}
\newcommand{\ee}{\end{equation}}

\begin{document}
\title{Galilean Exotic Planar Supersymmetries and Nonrelativistic Supersymmetric Wave Equations}
\author{J. Lukierski$^{1)}$, I. Pr\'ochnicka$^{1)}$,  P.C. Stichel$^{2)}$  and W.J. Zakrzewski$^{3)}$
\\
\siz $^{1)}$Institute for Theoretical Physics,  University of
Wroc{\l}aw, \\ \siz pl. Maxa Borna 9,
 50--205 Wroc{\l}aw, Poland\\
 \siz e-mail: lukier@ift.uni.wroc.pl; ipro@ift.uni.wroc.pl\\
\\\siz
$^{2)}$An der Krebskuhle 21, D-33619 Bielefeld, Germany \\ \siz
e-mail:peter@physik.uni-bielefeld.de
\\ \\ \siz
$^{3)}$Department of Mathematical Sciences, University of Durham, \\
\siz Durham DH1 3LE, UK \\ \siz
 e-mail: W.J.Zakrzewski@durham.ac.uk
 }

\date{}
\maketitle

\begin{abstract}
We describe the  general class of $N$-extended $D=(2+1)$ Galilean
supersymmetries obtained, respectively, from the $N$-extended
$D=3$ Poincar\'{e} superalgebras with maximal sets of central
charges. We confirm the consistency of supersymmetry with the
presence of the `exotic' second central charge $\theta$. We show
further how to introduce a $N=2$ Galilean superfield equation
describing nonrelativistic spin $0$ and spin $\frac{1}{2}$ free
particles.

\end{abstract}

\newpage

\section{Introduction}

The Galilean invariance is the fundamental space-time symmetry of nonrelativistic systems. The Galilean
projective representations (see e.g. \cite{1}), used in the description of quantum
$d$-dimensional nonrelativistic systems,
are generated by a central
extension of the $D=(d+1)$ Galilei algebra called also Bargmann or
quantum Galilei algebra, with mass $m$ as a central charge.
For example, for $d=3$ the Galilei algebra takes the following
form $(r,s,t=1,2,3$; we assume that the generators are Hermitean):
$$ [J_r,\,J_s]\,=\,i\,\epsilon_{rst}\,J_t,$$
$$ [J_r,\,K_s]\,=\,i\,\epsilon_{rst}\,K_t,$$
$$ [K_r,\,K_s]\,=\,0,\quad [P_r,\,P_s]\,=\,0,$$
$$ [J_r,\,P_s]\,=\,i\,\epsilon_{rst}\,P_t,$$
\begin{equation}
[H,\,J_s]\,=\,0,\quad [H,\,P_s]\,=\,0,
\label{oneone}
\end{equation}
$$ [H,\,K_r]\,=\,i\,P_r,$$
$$ [P_r,\,K_s]\,=\,i\,\delta_{rs}\,m,$$
where $J_r$ describe $O(3)$ rotations, $K_r$ - Galilean boosts, $P_r$ - momenta and
$H$ is the energy operator. It is known that the Galilei algebra (1) can be obtained
from the $D=4$ Poincar\'{e} algebra by the relativistic contraction $c\,\rightarrow\, \infty$ (\cite{2}).

The (2+1) dimensional Galilean algebra is a special case as, exceptionally, it allows for the
existence of a second central charge $\theta$. This algebra, with central charges $m$ and $\theta$,
also called `exotic', takes the following form (see e.g. \cite{3}-\cite{5}) $(i,j=1,2$)
$$ [J,\,K_i]\,=\,i\,\epsilon_{ij}\,K_j,$$
$$ [K_i,\,K_j]\,=\,i\,\epsilon_{ij}\,\theta,\qquad [P_i,\,P_j]\,=\,0,$$
\begin{equation}
[J,\,P_i]\,=\,i\,\epsilon_{ij}P_j,\quad [H,\,J]\,=\,[H,\,P_i]\,=\,0,
\label{onetwo}
\end{equation}
$$ [H,\,K_i]\,=\,i\,P_i,$$
$$ [P_i,\,K_j]\,=\,i\,\delta_{ij}\,m,$$
with the generator $J$ describing the $O(2)$ rotations.
In \cite{6} it was shown that the second central charge $\theta$ can be reinterpreted as introducing
noncommutativity in the two-dimensional space
\begin{equation}
[\hat x_i,\,\hat x_j]\,=\,-i\,\frac{\theta}{m^2}\,\epsilon_{ij}.
\label{onethree}
\end{equation}
The distinguished role of planar Galilean algebra follows from the
covariance of the relations (3) under space rotations. For $d>2$
($d$ - number of space dimensions) it is not possible to have
noncommutativity relation for  space coordinates which are
covariant under classical $O(d)$ rotations what corresponds to the
property that the respective  Galilean algebras do not allow for
the existence of the second central charge.

The first attempt at obtaining supersymmetric extensions of the $D=(3+1)$ Galilean algebra (\ref{oneone})
was made by Puzalowski (\cite{7}) who considered the $N=1$ and $N=2$ cases.
 The supersymmetric extension of the
$D=(2+1)$ Galilean symmetries were considered for $N=1$ and $N=2$ in \cite{8}-\cite{11}.
We should observe that the Galilean symmetries were extended to the so-called
Schr\"odinger symmetries of
free quantum mechanical systems (free nonrelativistic particle (\cite{12})
and harmonic oscillator (\cite{13}))
by adding two additional
generators: $D$ (dilatations) and $K$ (extensions, corresponding to the conformal time transformations).
The Galilean superalgebra was then obtained as a subalgebra of
supersymmetrically extended Schr\"odinger algebra (\cite{9}, \cite{14}-\cite{16}).
However, the $N$ extended supersymmetrization of the $D=(2+1)$
Galilean algebras  have, so far, not been described
in its general form; in particular,
the $D=(2+1)$ Schr\"odinger super-algebra has never been
written down in the presence of the second central charge $\theta$.

The aim of this paper is twofold:
\begin{itemize}
\item
To describe the new class of $N-$ extended (N even) $D=(2+1)$
 Galilean superalgebras, which we obtain
by the nonrelativistic contraction $c\rightarrow \infty$
($c$ - velocity of light) of the corresponding $D=3$
relativistic Poincar\'{e} superalgebras with maximal sets of central
 charges. We obtain
$\frac {N(N-1)}{2}$ Galilean central charges - {\it i.e.} the same
number as in the relativistic case\footnote{By central charges
we denote here the Abelian generators which commute with supercharges
 and Galilean generators. In general they
transform as tensors under the internal symmetry group.}. In Sect. 2. we show that
 the contraction of the central charges sector, which
preserves their number, requires a suitable rescaling to obtain finite
results in the $c \rightarrow \infty$ limit. We show further that for $D=(2+1)$ the two-parameter
($m,\theta)$ central extension (see (\ref{onetwo})) is consistent with supersymmetry.

Note that an alternative split of supercharges into two sectors
with different rescalings, by $\sqrt{c}$ and $\frac{1}{\sqrt{c}}$
factors, was earlier considered for the $D=10$  $N=2$ SUSY in
[17]. We shall adapt the method of [17] to $D=3$ and compare
the results with our contraction scheme.

\item
To present the realization of the $N=2$ $D=(2+1)$ Galilean superalgebra describing the supersymmetric
nonrelativistic particle multiplet with spin $0$ and $\frac{1}{2}$. In particular, we obtain
 the Levy-Leblond equations for nonrelativistic spin $\frac{1}{2}$
 fields (\cite{18}). The model can easily be made invariant under the
  exotic planar Galilean supersymmetry with the central charge $\theta\ne 0$.
\end{itemize}

Let us add that recently the Galilean supersymmetries have been
applied to the light-cone description of superstrings
(\cite{19}), nonrelativistic super-p-branes (\cite{20,21}) and D-branes (\cite{22}).
We conjecture that our $D=(2+1)$ Galilean supersymmetries with
central charge $\theta\ne0$ can find application in
 the description of nonrelativistic
supermembranes with noncommutative world volume geometry.

\section{$N$- extended Galilean $D=(2+1)$ superalgebras as contraction limits}

Let us recall that the $N$-extended $D=3$ Poincar\'{e} superalgebra
is given by ($\alpha,\beta=1,2$; $\mu=0,1,2$; $A,B=1..N$):
\begin{equation}
\{Q_{\alpha}^A,\,Q_{\beta}^B\}\,=\,(\sigma_{\mu})_{\alpha\beta}\,P^{\mu}\,\delta^{AB}\,+\,\epsilon_{\alpha\beta}Z^{AB},
\label{twoone}
\end{equation}
where $P^{\mu}=\eta^{\mu\nu}P_{\nu}$ ($\eta_{\mu\nu}=$diag(-1,1,1)), $Z^{AB}=-Z^{BA}$ are $\frac{N(N-1)}{2}$ real central charges and
$\sigma_{\mu}=(\sigma_i=\gamma_0\gamma_i,$ $\sigma_0=\gamma_0^2=-1_2$). We choose
\be
\label{twotwo}
\gamma_1=\left( \begin{array}{cc}
0&1\\1&0\end{array}\right),\quad \gamma_2=\left( \begin{array}{cc}
1&0\\0&-1\end{array}\right), \quad \gamma_0=\left( \begin{array}{cc}
0&1\\-1&0\end{array}\right)\,=\,\epsilon.
\ee

The full superalgebra is described by $2N$ real supercharges $Q_{\alpha}^A$, the $D=3$
Poincar\'{e} algebra ($P_{\mu}=(P_0,P_i),\quad M_{\mu\nu}=(M_{12}=J$ $,M_{i0}
=N_i)$),
$\frac{N(N-1)}{2}$ central charges $Z^{AB}$ and the $O(N)$
generators $T^{AB}=-T^{BA}$ describing internal symmetries:
\begin{itemize}
\item i)
$D=3$ Poincar\'{e} algebra
$$[J,\,N_i]\,=\,i\,\epsilon_{ij}N_j$$
$$[N_i,\,N_j]\,=\,i\,\epsilon_{ij}J$$
\be
\label{twothree}
[J,\,P_i]\,=\,i\,\epsilon_{ij}P_j,\quad [J,\,P_0]\,=\,0
\ee
$$[P_0,\,N_i]\,=\,i\,P_i,\quad [P_0,\,P_i]\,=\,[P_i,\,P_j]\,=\,0,$$
$$[P_i,\,N_j]\,=\,i\,\delta_{ij}P_0.$$
\item ii)
Supercovariance relations for $2N$ real supercharges $Q_{\alpha}^A$
$$[Q_{\alpha}^A,\,N_i]\,=\,\frac{i}{2}(\sigma_i)_{\alpha\beta}Q^{\beta A},\qquad [Q_{\alpha}^A,\,P_i]\,=\,0,$$
\be
\label{twofour}
[Q_{\alpha}^A,J]\,=\,\frac{i}{2}\epsilon_{\alpha\beta}Q^{\beta A},\qquad [Q_{\alpha}^A,\,P_0]\,=\,0.
\ee
\item iii)
The internal index $A$ is rotated by the $O(N)$ generators $T^{AB}$, where
$$[T^{AB},\,P^{\mu}]\,=\,[T^{AB},\,M^{\mu\nu}]\,=\,0,$$
\be
\label{twofive}
[T^{AB},\,T^{CD}]\,=\,i\,\left(\delta^{AC}T^{BD}\,-\,\delta^{AD}T^{BC}\,+\,\delta^{BD}T^{AC}\,-\,\delta^{BC}T^{AD}\right)
\ee
and
$$[T^{AB},\,Q_{\alpha}^C]\,=\,i\,(\tau^{AB})^C_D\,Q_{\alpha}^D,$$
where $(\tau^{ab})_c^d$ describe the vectorial $N\times N$ matrix  representation of $O(N)$ and
\be
\label{twosix}
[T^{AB},\,Z^{CD}]\,=\,i\,\left(\delta^{AC}Z^{BD}\,-\,\delta^{AD}Z^{BC}\,+\,\delta^{BD}Z^{AC}\,-\,\delta^{BC}Z^{AD}\right).
\ee
\item iv)
Central charges $Z^{AB}$ are Abelian and commute with the supercharges $Q_{\alpha}^A$ and with $(P_{\mu},M_{\mu\nu})$).
For a given choice $Z_{(0)}^{AB}$ of the central charges the internal unbroken
symmetry is described by the generators $\tilde T^{AB}\in T^{AB}$ which satisfy the relation (c.p. (\ref{twosix}))
\be
\label{twoseven}
[\tilde T^{AB},\,Z^{CD}_{(0)}]\,=\,0.
\ee
\end{itemize}

The nonrelativistic contraction of the $D=3$ Poincar\'{e} algebra
part of the super-algebra  (see (\ref{twothree})) is obtained by the
introduction of the  Hamiltonian $H$ in the following way:
\be
\label{twosevena}
P_0\,=\,mc\,+\,\frac{1}{c}H.
\ee
If we now perform the rescaling
$$ N_i\,=\,cL_i$$
and take the limit $c\rightarrow \infty$ then we obtain from (\ref{twothree})
the relations (\ref{onetwo}) with $\theta=0$ and $K_i$ replaced by $L_i$.
In order to get the algebra (\ref{onetwo}) ({\it i.e.} the exotic
(2+1) - dimensional Galilean algebra) one should perform the following linear
change of basis
\be
\label{twosevenb}
K_i\,=\,L_i\,+\,\frac{\theta}{2m}\,\epsilon_{ij}\,P_j.
\ee

In the case of a simple $N=1$ $D=(2+1)$ Galilean superalgebra
the relation (\ref{twoone}) takes the simple form
\be
\label{twoeight}
\{Q_{\alpha},\,Q_{\beta}\}\,=\,\delta_{\alpha\beta}P_0\,+\,(\sigma_i)_{\alpha\beta}P_i.
\ee
Next we introduce the rescaled supercharges
\be
\label{twonine}
S_{\alpha}\,=\,\frac{1}{\sqrt{c}}Q_\alpha
\ee
and using (\ref{twoeight}) find that in the $c\rightarrow \infty$ limit
\be
\label{twoten}
\{S_{\alpha},\,S_{\beta}\}\,=\,\delta_{\alpha\beta}m.
\ee
Further, from (7) and (14)
we get the following completion of the $N=1$ $D=3$ Galilean super-algebra:
$$ [K_i,\,S_{\alpha}]\,=\,0$$
\be
\label{twoeleven}
[J,\,S_{\alpha}]\,=\,-\frac{i}{2}\,\epsilon_{\alpha\beta}S^{\beta}.
\ee

Next we derive the $N$-extended $D=(2+1)$-Galilean superalgebra for $N=2k$ ($k=1,2,..$).
We start from the superalgebra (\ref{twoone}) where $A,B=1,2...2k$. We define
 (see also \cite{22b})
\be
\label{twofourteen}
Q_{\alpha}^{\pm a}\,=\,Q_{\alpha}^a\,\pm\,\epsilon_{\alpha\beta}Q_{\beta}^{k+a},
\ee
where in the formula (17) we consider $a=1,...k$.
Using (\ref{twoone}) and (\ref{twofourteen}) we get ($a,b=1...k)$)
\be
\label{twofifteen}
\{Q_{\alpha}^{+a},\,Q_{\beta}^{+b}\}\,=\,2\delta_{\alpha\beta}P_0\delta^{ab}\,+\,\epsilon_{\alpha\beta}(Z^{ab}+Z^{\tilde a\tilde b})\,
+\,\delta_{\alpha\beta}(Y^{ab}-\tilde Y^{ab}),
\ee
\be
\label{twosixteen}
\{Q_{\alpha}^{+a},\,Q_{\beta}^{-b}\}\,=\,2(\sigma_iP_i)_{\alpha\beta}\delta^{ab}\,+\,\epsilon_{\alpha\beta}(Z^{ab}-Z^{\tilde a\tilde b})\,
+\,\delta_{\alpha\beta}(Y^{ab}+\tilde Y^{ab}),
\ee
\be
\label{twoseventeen}
\{Q_{\alpha}^{-a},\,Q_{\beta}^{-b}\}\,=\,2\delta_{\alpha\beta}P_0\delta^{ab}\,+\,\epsilon_{\alpha\beta}(Z^{ab}+Z^{\tilde a\tilde b})\,
-\,\delta_{\alpha\beta}(Y^{ab}-\tilde Y^{ab}),
\ee
where the $2k\times 2k$ matrix of central charges $Z^{AB}$ is described by the four $k\times k$ matrices
$$Z^{ab},%
 \quad \tilde Z^{ab}\,=\,{ Z}^{a+k\,b+k},$$
\be
\label{twoeightteen}
Y^{ab}\,=\,{ Z}^{k+a\,b},\quad \tilde Y^{ab}\,=\,{ Z}^{a\,k+b},
\ee
satisfying the symmetry properties:
\be
\label{twonineteen}
Z^{ab}\,=\,-Z^{ba},\qquad \tilde Z^{ab}\,=\,-\tilde Z^{ba}
\ee
$$
Y^{ab}\,=\,-\tilde Y^{ba}.
$$
Finally, one gets
($Y^{(ab)}=\frac{1}{2}(Y^{ab}+Y^{ba}),Y^{[ab]}=\frac{1}{2}(Y^{ab}-Y^{ba}$))
\be
\label{twotwentyone}
\{Q_{\alpha}^{+a},\,Q_{\beta}^{+b}\}\,=\,2\delta_{\alpha\beta}(P_0\delta^{ab}\,+\,Y^{(ab)})\,+\,\epsilon_{\alpha\beta}(Z^{ab}+\tilde Z^{ab})
\ee
\be
\label{twotwentytwo}
\{Q_{\alpha}^{+a},\,Q_{\beta}^{-b}\}\,=\,2((\sigma_iP_i)_{\alpha\beta}\delta^{ab}\,+\,\delta_{\alpha\beta}Y^{[ab]})\,+\,\epsilon_{\alpha\beta}(Z^{ab}-\tilde Z^{ab})
\ee
\be
\label{twotwentythree}
\{Q_{\alpha}^{-a},\,Q_{\beta}^{-b}\}\,=\,2\delta_{\alpha\beta}(P_0\delta^{ab}\,-\,Y^{(ab)})\,+\,\epsilon_{\alpha\beta}(Z^{ab}+\tilde Z^{ab}).
\ee
Before taking the nonrelativistic limit we rescale the supercharges as follows:
\be
\label{twotwentyfour}
S_{\alpha}^a\,=\,\frac{1}{\sqrt{c}}Q_{\alpha}^{+a},\quad R_{\alpha}^a\,=\,\sqrt{c}Q_{\alpha}^{-a}.
\ee

The limit
$c\rightarrow \infty$ of the relations (\ref{twotwentyone}-\ref{twotwentythree}) exists if  we assume that
$$ Y^{(ab)}\,=\,\delta^{ab}mc\,+\,\frac{\tilde Y^{(ab)}}{c},$$
\be
\label{twotwentyfive}
 Z^{[ab]}\,+\,\tilde Z^{[ab]}\,=\,\frac{\tilde U^{[ab]}}{c}
\ee
$$Z^{[ab]}\,-\,\tilde Z^{[ab]}\,=\,U^{[ab]}$$
with central charges  $Y^{[ab]}$, $U^{[ab]}$, $\tilde Y^{(ab)}$ and
$\tilde U^{(ab)}$ having finite $c\rightarrow \infty$ limits.
Using (\ref{twotwentyfour}) and (\ref{twotwentyfive}) we see that in this limit
$$\{S_{\alpha}^a,\,S_{\beta}^b\}\,=\,4m\,\delta_{\alpha\beta}\,\delta^{ab},$$
\be
\label{twotwentysix}
\{S_{\alpha}^a,\,R_{\beta}^b\}\,=\,2(\sigma_iP_i)_{\alpha\beta}\,+\,\delta_{\alpha\beta}Y^{[ab]}\,+\,\epsilon_{\alpha\beta}U^{[ab]}
\ee
$$\{R_{\alpha}^a,\,R_{\beta}^b\}\,=\,2\delta_{\alpha\beta}(H\delta^{ab}-\tilde Y^{(ab)})\,+\,\epsilon_{\alpha\beta}\tilde U^{[ab]}.
$$
We see that the $N$-extended (2+1)-dimensional algebra ($N=2k)$ (\ref{twotwentysix})
 contains the following set of
central charges, with their number given in the second row,
$$ U^{[ab]},\phantom{aaa}\quad Y^{[ab]},\qquad \tilde U^{[ab]},\qquad \tilde Y^{(ab)}$$
\be
\label{twotwentyseven}
\frac{k(k-1)}{2},\quad \frac{k(k-1)}{2},\quad \frac{k(k-1)}{2},\quad \frac{k(k+1)}{2},
\ee
As $\frac{k(k+1)}{2}+3\frac{k(k-1)}{2}=\frac{N(N-1)}{2}$ $(N=2k)$,
 the maximal number of central charges for the $D=3$ Poincar\'{e} and $D=(2+1)$
 Galilean symmetry is the same.

The supercovariance relations (\ref{twofour}) take the following form
\be
\label{twotwentyeight}
[N_i,\,Q_{\alpha}^{\pm a}]\,=\,-\frac{i}{2}\,(\sigma_i)_{\alpha\beta}Q_{\beta}^{\mp a}
\ee
and after rescalings (17) and (26) one gets, in the $c\rightarrow \infty$ limit
\be
\label{twotwentynine}
[L_i,\,S_{\alpha}^a]\,=\,0,\quad [L_i,\,R_{\alpha}^a]\,=\,-\frac{i}{2}(\sigma_i)_{\alpha\beta}S_{\beta}^a
\ee
or, after performing the shift (12)
\be
\label{twothirty}
[K_i,\,S_{\alpha}^a]\,=\,0,\quad [K_i,\,R_{\alpha}^a]\,=\,-\frac{i}{2}(\sigma_i)_{\alpha\beta}S_{\beta}^a.
\ee
Further, from (\ref{twofour}) we get
\be
\label{twothirtyone}
[J,\,S_{\alpha}^a]\,=\,-\frac{i}{2}\,\epsilon_{\alpha\beta}S_{\beta}^a,\quad [J,\,R_{\alpha}^a]\,=\,-\frac{i}{2}\,\epsilon_{\alpha\beta}R_{\beta}^a.
\ee

It is easy to see that the relations (\ref{twotwentysix}) are covariant under the $O(\frac{N}{2})=O(k)$ rotations, with indices $a,b$ describing the
$k$-dimensional vector indices.
In comparison with (\ref{twoone}) we see that the $O(N)$ covariance
of the relations (4) has been reduced in the contraction procedure to the covariance with
respect to the diagonal $O(k)$ symmetry obtained by
constraining the $O(N)$ generators $T^{AB}$ as follows:
\be
\label{aaaa}
T^{ab}\,=\,T^{k+a\,k+b},\,\qquad  T^{a\,k+b}\,=\,0.
\ee
 In consequence, all the central charges (\ref{twoeightteen}) are the second rank
$O(k)$ tensors and the Galilean supercharges ($S^a_{\alpha}$, $R^a_{\alpha}$) are the $O(k)$ vectors.

A special case corresponds to $k=1$ $(N=2)$, when ${ Z}^{AB}=\epsilon^{AB}{ Z}$
defines the scalar central charge $Z$ (from (22) we see
that for $k=1$,  $Y=-\tilde Y={ Z}$). One gets
$$ \{Q_{\alpha}^+,\,Q_{\beta}^+\}\,=\,2\delta_{\alpha\beta}(P_0+{ Z})$$
\be
\label{twothirtytwo}
\{Q_{\alpha}^+,\,Q_{\beta}^-\}\,=\,2(\sigma_iP_i)_{\alpha\beta}
\ee
$$ \{Q_{\alpha}^-,\,Q_{\beta}^-\}\,=\,2\delta_{\alpha\beta}(P_0-{ Z}).$$
Writing the first relation in (28) in the form
${ Z}=mc+\frac{U}{c}$ and using (11)  and (27) we get, when $c\rightarrow \infty$
$$ \{S_{\alpha},\,S_{\beta}\}\,=\,4m\delta_{\alpha\beta}$$
\be
\label{twothirtythree}
 \{S_{\alpha},\,R_{\beta}\}\,=\,2(\sigma_iP_i)_{\alpha\beta}
\ee
$$ \{R_{\alpha},\,R_{\beta}\}\,=\,2(H-U)\delta_{\alpha\beta},$$
where $U$ plays the role of the correction to the Hamiltonian originating
from the central charge $ Z$. Further, from the relation (8) we get $(T^{AB}=\epsilon^{AB}T;
A,B=1,2)$
\be
[T,\,Q_{\alpha}^{\pm}]\,=\,\mp i\,Q_{\alpha}^{\pm}
\ee
and after contraction
\be
[T,\,S_{\alpha}]\,=\,-i\,S_{\alpha},\quad [T,\,R_{\alpha}]\,=\,i\,R_{\alpha}
\ee
In addition we get, using (12), (27) and the contraction limit
\be
[K_i,\,S_{\alpha}]\,=\,0,\quad [K_i,\,R_{\alpha}]\,=\,-\frac{i}{2}(\sigma_i)_{\alpha\beta}S_{\beta}
\ee
as well as
\be
[J,\,S_{\alpha}]\,=\,-\frac{i}{2}\,\epsilon_{\alpha\beta}S_{\beta},\quad [J,\,R_{\alpha}]\,=\,-\frac{i}{2}\,\epsilon_{\alpha\beta}R_{\beta}.
\ee

In our contraction scheme we employ the split (17) of the
supercharges with two sectors undergoing different rescalings. If
we express the relations (17) as ($a=1,2$, $\alpha,\beta=1 \ldots
k$; N=2k)
\begin{equation}
\label{eq41} Q^{\pm a}_{\alpha}= ({\cal P}^{\pm})^{a \, b}_{\alpha
\, \beta} \, Q^{b}_{\beta}\, ,
\end{equation}
then the $N\times N$ matrices ${\cal P}^{\pm}$ do not describe the
projection operators\footnote{Because ${\cal P}^{\pm} =
\pmatrix{I_k & 0\cr 0 & \epsilon I_k}$ we get $$ {\cal P}_{\pm} \,
{\cal P}_{\pm} = \pmatrix{I_k & 0\cr 0 & - I_k}; \quad {\cal
P}_{\pm}{\cal P}_{\mp}= I_N\, .
$$}.
For $k=1$ ($N=2$) by adapting to $D=3$ the contraction scheme
proposed in [17] we can, however, introduce an alternative split of
the supercharges $Q^{a}_{\alpha}$
\begin{equation}
\label{eq42} \widetilde{Q}^{\pm a}_{\alpha} = (P_{\pm}
Q)_{\alpha}= Q^{a}_{\alpha} \pm \epsilon _{\alpha \beta} \epsilon
^{a b} Q^{b}_{\beta}\, ,
\end{equation}
where $P_{\pm}$ are the projection operators satisfying additional
properties
\begin{equation}
\label{eq43} P^{T}_{\pm} = P_{\pm} \qquad (P_{\pm})^{a b}_{\alpha
\beta} = \epsilon _{\alpha \gamma} \epsilon^{a c} \, P^{c
b}_{\gamma \beta} \, .
\end{equation}
As for $N=2$ we have $Z_{AB}= \epsilon_{AB}\ Z(A,B=1,2)$ we see
 from (4) that
\begin{eqnarray}
\label{eq44} \{ \widetilde{Q}^{\pm a}_{\alpha}, Q^{\pm
b}_{\beta}\} &= & 2 (P_{\pm})^{a b}_{\alpha \beta} (P_0 \pm Z)\, ,
\cr \{ {Q}^{+  a}_{\alpha}, Q^{- b}_{\beta}\} &= & 2 (P_{+})^{a
b}_{\alpha \gamma} (\sigma_i P_i)_{\gamma \beta}\, .
\end{eqnarray}
Introducing, in analogy with (26), the rescaling

\begin{equation}
\label{eq45} \widetilde{S}^{ a}_{\alpha} = \frac{1}{\sqrt{c}} \
\widetilde{Q}^{+ a}_{\alpha}\, , \qquad \widetilde{R}^{
a}_{\alpha} = {\sqrt{c}} \ \widetilde{Q}^{- a}_{\alpha}\, ,
\end{equation}
and using the relations (11) and (27) we get in the limit $c \to
\infty$ the following $N=2$  $\ D=3$ nonrelativistic Galilean
superalgebra

\begin{eqnarray}
\label{eq46} \{ \widetilde{S}^{ a}_{\alpha}, \widetilde{S}^{
b}_{\beta}\} &= & 4m (P_+)^{a b}_{\alpha \beta}\, , \cr \{
\widetilde{S}^{ a}_{\alpha}, \widetilde{R}^{ b}_{\beta}\} &= &
2(P_+)^{a b}_{\alpha \beta} (\sigma_i P_i)_{\gamma \beta}\, , \cr
\{ \widetilde{R}^{ a}_{\alpha}, \widetilde{R}^{ b}_{\beta}\} &= &
2(P_ -)^{a b}_{\alpha \beta} (H-U) \, .
\end{eqnarray}
The $D=10$ analogue of the superalgebras (44) and (46), under
a simplifying assumption $U=-H$, has been proposed in [17].
It appears that the projections (42) and the Galilean superalgebra (46) can be
used in supersymmetric
  $D$-brane models for the introduction of nonrelativistic kappa-symmetries
which eliminate half
of the fermionic degrees of freedom.

Finally we add that if $k>1$ the definition (42) and the appearance of kappa
transformations which eliminate half of the fermionic degrees of freedom
 remain valid
 in the presence of one nonvanishing central charge. In the presence
of several relativistic central charges the split of supercharges into the parts
undergoing $\sqrt{c}$ and $\frac{1}{\sqrt{c}}$ rescalings is not unique,
and the kappa transformations
 eliminate less than half of the fermionic variables.

\section{Superfield wave equation with standard and exotic $N=2$ Galilean supersymmetry}

In this section we derive a super-Galilean covariant wave equation for our
superfield $\Psi(t,x;\theta,\eta)$ where $\theta$ and $\eta$
 two  real valued anticommuting spinors with components $\theta_{\alpha}$ and $\eta_{\beta}$.

The proposed superfield equation is the following (see also \cite{28}):
\begin{equation}
\nabla_{\alpha} \Psi\,=\,0,
\end{equation}
where $\nabla_{\alpha}$ is a supercovariant derivative\footnote{Note that the supercovariance
 of this superderivative
holds only in a weak sense, {\it ie} as is clear from (56,57), it
is valid only on the solutions of the wave equation (47)}.

Here, we will describe a derivation of (47) without the
need of the Lagrangian.

In order to determine the explicit form of our supercovariant
derivative we require, in addition to (47) that we have
\begin{equation}
[\nabla_{\alpha},\,H]\,\Psi\,=\,0,\quad [\nabla_{\alpha},\,P_i]\,\Psi\,=\,0,
\end{equation}
\begin{equation}
\{\nabla_{\alpha},\,S_{\beta}\}\,\Psi\,=\,0,\quad \{\nabla_{\alpha},\,R_{\beta}\}\,\Psi\,=\,0,
\end{equation}
and
\begin{equation}
[\nabla_{\alpha},\,J]\,\Psi\,=\,0 \, ,\quad
[\nabla_{\alpha},\,K_{i}]\,\Psi\,=\,0.
\end{equation}

In addition (47) ought to be the ``square-root'' of the
Schr\"odinger equation [15]; {\it i.e.} we require that
\begin{equation}
\{\nabla_{\alpha},\,\nabla_{\alpha}\}\,\Psi\,\sim\,(H\,-\,\frac{\vec p^2}{2m})\,
\Psi.
\end{equation}

The super-derivative $\nabla_{\alpha}$, satisfying all the
requirements (48-51) is given by
\begin{equation}
\nabla_{\alpha}\,=\,\frac{1}{2}\,(\sigma_iP_i)_{\alpha\beta}\,S_{\beta}\,-\,m\,R_{\alpha}.
\end{equation}

To prove our claim we use the $N=2$ super-Galilean algebra with a vanishing
central charge $U$ discussed in section 2.

First we note that the invariance of $\nabla_{\alpha}$ with respect to
space-time translations is evident.
Furthermore,
\begin{enumerate}
\item from (36) we have
\begin{equation}
\{\nabla_{\alpha}, \nabla_{\alpha}\}\, =\, 4m^2(H-\frac{\vec P^2}{2m})
\end{equation}
\item and \begin{equation}
\{\nabla_{\alpha},\,S_{\beta}\}\,=\,0
\end{equation}
\item
from (39) and (40)
\begin{equation}
\qquad [\nabla_{\alpha},\,J]\,=\,\frac{1}{2}(\sigma_i)_{\alpha\beta}\nabla_\beta \,\, ,\quad [\nabla_{\alpha},\,K_i]\,=\,0
\end{equation}
\item
from (36)
\begin{equation} \qquad  \{\nabla_{\alpha},\,R_{\beta}\}\,=\,2m\,\delta_{\alpha\beta}\,(\frac{\vec P^2}{2m}-H)
\end{equation}
\end{enumerate}
and therefore, due to (47) and (53)
\begin{equation}
\{\nabla_{\alpha},\,R_{\beta}\}\Psi\,=\,0.
\end{equation}

In order to express the wave equation (47) with $\nabla_{\alpha}$
given by (52), in terms of derivatives with respect to the
arguments of $\Psi$, we need a realisation of the $N=2$
super-Galilei algebra in terms of differential operators. It is
easily seen that such a realisation is given by ($U=0$):
\begin{equation}
H\,=\,i\partial_t,\quad P_i\,=\,-i\partial_i,
\end{equation}
\begin{equation}
K_i\,=\,tP_i\,-mx_i\,+\,\frac{\theta}{2m}\epsilon_{ij}P_j\,+\,\frac{i}{2}\eta_{\alpha}(\sigma_i)_{\alpha\beta}\frac{\partial}{\partial \theta_\beta},
\end{equation}
\begin{equation}
J\,=\,\epsilon_{ij}x_iP_j\,+\,\frac{i}{2}\epsilon_{\alpha\beta}\left(\theta_{\alpha}
\frac{\partial}{\partial \theta_{\beta}}\,+\,\eta_{\alpha}\frac{\partial}{\partial \eta_{\beta}}\right),
\end{equation}
\begin{equation}
S_{\alpha}\,=\,2\frac{\partial}{\partial\theta_{\alpha}} \, +\,\frac{1}{2}
(\sigma_iP_i)_{\alpha\beta}\eta_{\beta}\,+\,m\theta_{\alpha},\end{equation}
\begin{equation}
R_{\alpha}\,=\,2\frac{\partial}{\partial\eta_{\alpha}} \, +\,\frac{1}{2}
(\sigma_iP_i)_{\alpha\beta}\theta_{\beta}\,+\,\frac{i}{2}
\eta_{\alpha}\partial_t,\end{equation}
where the spinor derivatives are defined as left-derivatives and
the $x_i$ are commuting
variables.

Note that the spinor part of $K_i$ is necessary so that $K_i$ has the correct
expressions for its commutators with the supercharges $S_{\alpha}$
and
$R_{\alpha}$. This extra term leads also to a nontrivial behaviour of the
spinor $\theta_{\alpha}$ with respect to boosts {\it i.e}:
\begin{equation}
[K_i,\,\theta_{\alpha}]\,=\,\frac{i}{2}(\sigma_i)_{\alpha\beta}\eta_{\beta}.\end{equation}

{}From (61) and (62) we can read off the transformation properties
of our variables in superspace $Y\in(t,x,\theta,\eta)$ under
infinitesimal supertranslations. Then with
\begin{equation}
\delta Y\,=\,[Y,\,\epsilon_{\beta}S_\beta+\rho_\beta R_\beta]
\end{equation}
we obtain
\begin{equation}
\delta \eta_{\alpha}\,=\,2\rho_{\alpha},\quad \delta \theta_{\alpha}\,=\,2\epsilon_{\alpha},
\end{equation}
\begin{equation}
\delta x_i\,=\, \frac{i}{2}(\epsilon_{\beta}(\sigma_i)_{\beta\gamma}\eta_{\gamma}\,+\,
\rho_{\beta}(\sigma_i)_{\beta\gamma}\theta_{\gamma}),
\end{equation}
\begin{equation}
\delta t\,=\,-\frac{i}{2}\,\rho_{\beta}\eta_{\beta}.
\end{equation}

Finally, using (61) and (62) for the supercharges $S_{\alpha}$,
$R_{\alpha}$ we obtain the following decomposition of the
super-covariant derivative $\nabla_{\alpha}$
\begin{equation}
\nabla_{\alpha}\,=\,D_{\alpha}\,+\,\frac{1}{4}\eta_{\alpha}(\vec P^2\,-\,2mi\partial_t),\end{equation}
where
\begin{equation}\label{spin}
D_{\alpha}\,=\,(\sigma_iP_i)_{\alpha\beta}\frac{\partial}{\partial \theta_{\beta}}\,-\,2m\frac{\partial}{\partial\eta_{\alpha}}.\end{equation}

Thus the wave equation (47) is equivalent to the following pair of
differential equations:
\begin{equation}
\label{bbbb}
D_{\alpha}\Psi\,=\,0,
\end{equation}
with $D_{\alpha}$ given by (69) and
\begin{equation} \label{aa}
\left(i\partial_t\,-\,\frac{\vec P^2}{2m}\right)\Psi\,=\,0.\end{equation}

As the superfield $\psi$ has
 the expansion
\begin{equation}
\Psi(t,x;\theta_{\alpha},\eta_{\alpha})\,=\,\phi(t,x)\,+\,\theta_{\alpha}\psi_{\alpha}(t,x)
\,+\,\eta_{\alpha}\chi_{\alpha}(t,x)\,+...
\end{equation}
the equation (\ref{bbbb}) gives us the following equation  for the spinor
fields
\begin{equation}
(\sigma_iP_i)_{\alpha\beta}\psi_{\beta}(t,x)
\,-\, 2m\chi_{\alpha}(t,x)\,=\,0\end{equation}
which, when combined with (\ref{aa}) gives us
\begin{equation}
i\partial_t\psi_{\alpha}\,=\,\frac{1}{2}(\sigma_iP_i)^2\psi_{\alpha}
\,=\,(\sigma_iP_i)_{\alpha\beta}\chi_{\beta}.\end{equation} The
set of Eqs.(73)-(74) provides the Levy-Leblond equations for the
nonrelativistic spin 1/2 particles ([18]; see also [8]).

\section{Conclusions}

In this paper we have restricted ourselves to the case of the $D=2+1$ nonrelativistic
supersymmetries but the discussion of the $D=3+1$ case can be
performed in an analogous way. In particular, the basic equations
 (52), (70) and (73)
 can be extended to  a three-dimensional space by introducing complex two-component spinors
 $\theta_\alpha$, $\eta_\alpha$ and three 2x2 complex Pauli matrices.

Note that in Section 3 we derived the superfield form of the
nonrelativistic Levy-Leblond equations without postulating a classical action in
the $N=2$ $\ D=3$ superspace $(x_{i}, t , \theta_\alpha ,\eta_\alpha)$.
Note also that our superfield equation (see (47) and (68))
  does not depend on the exotic parameter
$\theta$.

In Section 2 we have presented the new general nonrelativistic
contraction scheme for the $N$-extended $D=3$  Poincar\'{e} algebra.
 Had we started from the most general $D=4$ N-extended Poinca'{e} algebra
 with $\frac{N(N-1)}{2}$ complex central charges (\cite{29}), we would have obtained
 analogous results for the $D=3+1$ nonrelativistic supersymmetric theory,
  with equal numbers of relativistic and nonrelativistic complex central charges.

\subsection*{Acknowledgments}
Two of the authors (JL) and (PCS) would like to thank the
University of Durham for hospitality and the EPSRC for financial support.
The authors would also like to thank the referee, who pointed out to them
the relevance of ref. \cite{21}.

 Partial financial support by a KBN grant 1 P03B 01828 is also
acknowledged (J.L.).

\end{document}